\documentclass{JHEP3}

\usepackage{amsmath}
\usepackage{amssymb}
\usepackage{graphicx}
\usepackage{slashed}
\usepackage{multirow}

\newcommand{\cN}{\mathcal{N}}

\newcommand{\ov}{\overline}

\newcommand{\drawsquare}[2]{\hbox{%
\rule{#2pt}{#1pt}\hskip-#2pt
\rule{#1pt}{#2pt}\hskip-#1pt
\rule[#1pt]{#1pt}{#2pt}}\rule[#1pt]{#2pt}{#2pt}\hskip-#2pt
\rule{#2pt}{#1pt}}
\newcommand{\Ysymm}{\raisebox{-.5pt}{\drawsquare{6.5}{0.4}}\hskip-0.4pt%
         \raisebox{-.5pt}{\drawsquare{6.5}{0.4}}}
\newcommand{\Yasymm}{\raisebox{-3.5pt}{\drawsquare{6.5}{0.4}}\hskip-6.9pt%
        \raisebox{3pt}{\drawsquare{6.5}{0.4}}}

\title{Singlet Extensions of the MSSM in the Quiver Landscape}

\author{Mirjam Cveti\v{c}$^{1,2,4}$, James Halverson$^{1,4}$, and Paul Langacker$^{3,4}$\\
  $^1$Department of Physics and Astronomy, University of Pennsylvania,\\
  Philadelphia, PA 19104-6396, USA\vspace{.5cm} \\
  $^2$Center for Applied Mathematics and Theoretical Physics,\\
University of Maribor, Maribor, Slovenia\vspace{.5cm} \\
  $^3$School of Natural Science, Institute for Advanced Study,\\
  Einstein Drive, Princeton, NJ 08540, USA \vspace{.5cm}\\
  $^4$Kavli Institute for Theoretical Physics, Kohn Hall,\\
  UCSB, Santa Barbara, CA 93106, USA \vspace{.5cm}\\
  E-mail: \email{cvetic@cvetic.hep.upenn.edu}, \email{jhal@physics.upenn.edu}, \email{pgl@ias.edu}}

\abstract{We map out possible extensions of the MSSM in the context of type II string
theory. We systematically investigate three-stack and four-stack quivers which realize the MSSM 
spectrum with the addition of a single MSSM singlet $S$ with an allowed $S H_u H_d$ term,
which can lead to a dynamical electroweak-scale $\mu$-term.
We present the three quivers which satisfy stringent string-theoretic and phenomenological
constraints, including the presence of non-zero masses for all three families of quarks and leptons,
 the perturbative and non-perturbative absence of R-parity violating couplings and rapid dimension-five proton decay, and a mechanism for small neutrino masses. We find that these quivers can realize many models in the class of singlet-extended (supersymmetric)
standard models, as D-instanton effects can in principle
generate a superpotential of the form $f(S)$, where $f$ is a polynomial.
Finally, we address the issue of the stabilization and decoupling of charged moduli which generically
appear in D-instanton corrections to the superpotential.}

\preprint{UPR-1217-T\\NSF-KITP-10-087}

\begin{document}

\section{Introduction}
It has long been recognized that string theory naturally realizes important ingredients seen in
particle physics, including gauge symmetry and chiral matter. More recently, the field has progressed
tremendously and it is now possible to investigate stringy effects which could give rise to the
finer details seen in particle physics. D-brane instantons \cite{Blumenhagen:2006xt,Florea:2006si,Ibanez:2006da,Blumenhagen:2009qh} have been particularly fruitful in this regard,
as they give non-perturbative superpotential corrections which could explain the large mass hierarchies
\cite{Anastasopoulos:2009mr,Cvetic:2009ez,Cvetic:2009ng}
and mixing angles seen in the standard model, in addition to possibly playing a role in supersymmetry
breaking \cite{Cvetic:2007qj,Cvetic:2008fj}, and moduli stabilization \cite{Kachru:2003aw,Balasubramanian:2005zx,Blumenhagen:2007sm}.
Furthermore, they often generate operators relevant for obtaining small neutrino masses, including
a Majorana mass term (\cite{Blumenhagen:2006xt}, \cite{Ibanez:2006da}, \cite{Cvetic:2007ku,Ibanez:2007rs,Antusch:2007jd}),
a small Dirac mass term \cite{Cvetic:2008hi}, or a stringy Weinberg operator \cite{Cvetic:2010qy}.

With this progress in our understanding of string vacua and the start-up of the LHC, 
it is important to consider the possible implications of string theory for particle physics
in as much detail as possible. The natural starting place is to consider models which realize the
exact MSSM, possibly extended by three right-handed neutrinos,  and much work has been conducted
along these lines. However, it is also important to consider the question of what
allowed extensions of the standard model or MSSM are likely to occur in the string landscape.
The
minimal MSSM spectrum is not
necessarily phenomenologically preferred, as the MSSM has some fine-tuning problems of its own,
especially the $\mu$-problem.
Furthermore, string constructions frequently contain extra chiral matter beyond
the MSSM spectrum, which are often MSSM singlets.  
There is an extensive literature of phenomenological studies of such singlets, especially those
which allow for a dynamical solution to the $\mu$ problem\footnote{See, e.g.,  \cite{Ellis:1988er,Suematsu:1994qm,Cvetic:1995rj,Panagiotakopoulos:1999ah,Panagiotakopoulos:2000wp,Dedes:2000jp,Menon:2004wv}. For recent reviews, see \cite{Accomando:2006ga,Barger:2007ay,Maniatis:2009re,Ellwanger:2009dp}.}, but relatively little study
from the string perspective\footnote{See \cite{Chemtob:2009ie,Lebedev:2009ag} for recent studies in heterotic constructions.}.

Rather than considering full string models, we work from the bottom-up \cite{Aldazabal:2000sa,Antoniadis:2000ena,Antoniadis:2001np}  at the level
of quivers \cite{Ibanez:2008my,Leontaris:2009ci,Anastasopoulos:2009mr,Cvetic:2009yh,Cvetic:2009ez,Kiritsis:2009sf,Cvetic:2009ng,Anastasopoulos:2010ca}, which allow for the efficient classification of many important physical effects. The data associated
with a quiver essentially amounts to  the gauge symmetry and matter content, which are data naturally associated
with any semi-realistic string model. However, one can also examine the possible presence or absence
of couplings and other physical effects by examining quantum numbers associated with the quiver, without
having to delve into the geometric specifics of a fully defined string model. In this sense, we probe the ``quiver landscape" for physical effects,
hoping to identify promising quivers which could arise at various points in the landscape of string vacua.

Such an investigation was carried out for the exact MSSM and its extension by three right-handed neutrinos in
\cite{Cvetic:2009yh,Cvetic:2010qy}. Around fifty quivers were found with this matter spectrum which satisfy necessary constraints for the
cancellation of non-abelian anomalies and a massless hypercharge, as well as allowing for Yukawa couplings for all three
families of quarks and leptons via D-instanton effects. In those quivers, instantons whose presence is required to induce
Yukawa couplings do not give rise to phenomenological drawbacks, such as violation of R-parity, rapid dimension-five 
proton decay, or a $\mu$-term which is too large. Additionally, each 
admits a mechanism 
which can give rise to neutrino masses of the correct order, and many contain semi-realistic mass hierarchies
for the quarks and leptons.

In this work we perform an analysis of quivers whose spectrum is given by the exact MSSM extended by a singlet $S$, as
well as quivers where this spectrum is further extended by the addition of three right-handed neutrinos.
Models with such a spectrum belong to the class
of so-called singlet-extended standard models. Such extensions are phenomenologically
motivated as a solution to $\mu$-problem, where the $\mu$-term can be dynamically generated by a coupling $SH_uH_d$
when the $S$ field obtains a vacuum expectation value. In addition, singlet extensions are motivated from string
theory by the simple fact that one or more standard model singlets often appear in the massless spectrum. 

We present the three singlet-extended quivers with less than five stacks which satisfy a 
host of string theoretic and phenomenological constraints, including those mentioned above. 
We consider possible instanton corrections to the superpotential, which
would determine the particular singlet-extended standard model that the quiver realizes, and find that
the superpotential corrections are generically a polynomial function $f(S)$. Thus, the quivers themselves
are \emph{model independent} and could give rise to a variety of particular singlet-extended models
when embedded in a global string model. The determination of the particular singlet-extended model
associated with a global model depends heavily on the geometry of the Calabi-Yau, and we leave
this analysis for future work.

We also address the issue of the stabilization of charged moduli which arise in instanton corrections to the superpotential.
These moduli transform under the gauge symmetry of the D-brane which the D-instanton intersects. 
Assuming that the uncharged moduli are
stabilized at high scale in some hidden sector where SUSY is broken, we examine the $F$-term and $D$-term contributions
to the scalar potential in detail. Minimization of the potential shows that the real part of the charged moduli stabilizes near zero VEV, fixes
the $D$-term near zero, and decouples from the low energy effective action. 

This paper is organized as follows. In section \ref{sec:type II}, we review that basics of quivers and D-instantons
in type II orientifold compactifications and discuss what can be said about non-perturbative corrections at
the level of a quiver. In section \ref{sec:beyond}, we discuss some basic motivations for singlet-extended
standard models and discuss the role of instantons in the non-perturbative generation of crucial superpotential
terms. We also present three quivers which satisfy many string theoretic and phenomenological constraints,
discussing possible superpotential corrections 
that determine the
type of singlet-extended standard model and neutrino masses. In appendix \ref{sec:quiver appendix},
we explicitly present the constraints which the three quivers satisfy and discuss the methodology of our
systematic analysis. Finally, in appendix \ref{sec:moduli appendix}, we show that the real part of the charged moduli associated
with instanton corrections generically stabilize near zero VEV and decouple from the low energy effective
action.

\section{Quivers in Type II Orientifold Compactifications \label{sec:type II}}
The quivers analyzed and presented in this work fit naturally into orientifold compactifications of type II
string theory \cite{Blumenhagen:2005mu,Blumenhagen:2006ci,Marchesano:2007de}\footnote{For examples of
globally consistent supersymmetric standard-like models, see \cite{Cvetic:2001nr,Cvetic:2001tj}.}.
For the sake of concreteness we consider type IIa, where gauge theories live on D6-branes which wrap four-dimensional
Minkowski space and a three-cycle in the internal Calabi-Yau threefold. The D-branes carry Ramond-Ramond
charge, which must be cancelled by the introducton of orientifold planes, whose locations are given by the fixed
point loci of an antiholomorphic involution on the internal space. The cancellation of Ramond-Ramond charge, also
known as tadpole cancellation, is a condition on the homology of the D6-branes and O6-planes.

D6-branes naturally give rise to gauge symmetry and chiral matter. When wrapped on a generic cycle, a stack of $N$
D6-branes gives rise to $U(N)=SU(N)\times U(1)$ gauge symmetry. Additionally, $N$ D6-branes on cycles which are homologically-fixed
or pointwise-fixed by the orientifold action give rise to $Sp(2N)$ and $SO(2N)$ gauge symmetry, respectively. Chiral matter arises
in the bifundamental representation at the non-trivial intersection of two generic D6-branes, and it is also possible
to have symmetric or anti-symmetric tensor representations where a D6-brane intersects its image brane under
the orientifold action. Furthermore, two given D6-branes might intersect in multiple points on the compact internal space,
giving rise to multiple families, where the number of families is the topological intersection number of the two D6-branes
in the Calabi-Yau.

Of great phenomenological importance is the generalized Green-Schwarz mechanism, which generically gives a St\" uckelberg
mass to the gauge bosons associated with the $U(1)$ symmetries of the D6-branes, but might leave some linear combination
massless, depending on the homology of the D-branes and O-planes. The presence of such a massless $U(1)$ is crucial in type II model building, as it can be interpreted as hypercharge. Additionally, the $U(1)$ combinations which do receive a mass
still exist as global symmetries of the theory, which charge chiral matter and generically forbid many Yukawa couplings
at the perturbative level in the superpotential. 
In the absence of non-perturbative corrections, this would yield some number of massless quark and lepton families.


However, in \cite{Blumenhagen:2006xt} it was shown that Euclidean D2-instantons,
which are pointlike in spacetime and wrap a three-cycle in the Calabi-Yau, can non-perturbatively generate
a superpotential coupling of the generic form
\begin{equation}
W \sim e^{-S_{E2}^{cl}} \prod_i \Phi_i,
\end{equation}
where the $\Phi_i$ are matter fields charged under the gauge group of some D6-branes. In addition to generating
couplings which might be perturbatively forbidden, it has been shown that such
$E2$-instantons can generate phenomenologically relevant couplings which are always perturbatively forbidden in type II, 
such as the $10\,10\,5_H$ top-quark Yukawa coupling in $SU(5)$ GUT models \cite{Blumenhagen:2008yq} or a Majorana mass term for right-handed neutrinos
(\cite{Blumenhagen:2006xt}, \cite{Ibanez:2006da}, \cite{Cvetic:2007ku,Ibanez:2007rs,Antusch:2007jd}).
Furthermore, D-instantons can also account for the observed order of the neutrino masses by inducing \cite{Cvetic:2010qy} a Weinberg operator \cite{Weinberg:1980bf} directly (for a lower string scale) or via a highly suppressed Dirac mass term \cite{Cvetic:2008hi}.

Determining the form of instanton-induced superpotential corrections, if any, requires a careful analysis of the
uncharged and charged fermionic zero modes, respectively in the E2-E2 and E2-D6 open string sectors. In addition to
the $x^\mu$ and $\theta^\alpha$ modes which are present in the measure of the superpotential, the uncharged modes include
deformation modes and a $\ov\tau_{\dot \alpha}$ mode which must be projected out or lifted in order for the
instanton to generate a superpotential contribution. This occurs, for example, when the instanton wraps a rigid
orientifold invariant cycle, which is a condition that depends heavily on geometric specifics. Thus, it is not
possible to address these zero modes merely from a quiver analysis.

The charged zero modes, on the other hand, are charged under the gauge groups of the D6-branes and thus have
quantum numbers which can be seen at the quiver level. For example, consider the simple case of three $U(1)$ stacks
of D6-branes $a$, $b$, and $c$, with fields $\Phi$ and $\Psi$ which transform as $(a,\ov{b})$ and $(b,\ov c)$.
In that case the coupling
\begin{equation}
\Phi_{(1,-1,0)}\Psi_{(0,1,-1)}
\end{equation}
has non-zero charge under the global $U(1)$'s, which are denoted by subscripts, and is therefore perturbatively forbidden. Any instanton which might generate
the coupling non-perturbatively must cancel the $U(1)$ charge, and thus must give rise to a $\ov{\lambda}_a$ and a
$\lambda_c$ charged mode. In this way, it is straightforward to determine not only which charged modes must be present
if an instanton is to generate a particular superpotential coupling, but also which superpotential couplings might
be generated by the same instanton. These necessary conditions can all be seen at the quiver level, without having to specify the geometry of a particular Calabi-Yau.



\section{Beyond the MSSM}
\label{sec:beyond}
Since the quiver approach allows for the investigation of many physical effects without the necessity of specifying
a particular string geometry, it is worthwhile to examine quivers which realize promising phenomenological models.
The natural starting point is with quivers which realize the exact spectrum of the MSSM, 
possibly extended by three right-handed neutrinos.
A systematic investigation of MSSM quivers at the level of gauge symmetry and matter content
was performed in \cite{Anastasopoulos:2006da}, and a systematic analysis of realistic MSSM quivers was first
performed at the level of couplings in \cite{Cvetic:2009yh,Cvetic:2010qy}, taking into account non-perturbative D-instanton
effects and extensive phenomenological considerations. 

While those MSSM quivers are phenomenologically promising and provide an excellent starting place for string model-building, 
there are motivations for
adding additional chiral matter to the spectrum. For example, the MSSM
superpotential contains a term $\mu H_u H_d$. 
The $\mu$ parameter must be non-zero and above $\sim 100$ GeV to ensure sufficiently large Higgsino masses,
while the associated soft term must be non-zero to  ensure that both $H_u$ and $H_d$ obtain
nonzero vacuum expectation values after electroweak symmetry breaking. Both should be
 below the TeV scale to allow a solution to the Higgs hierarchy problem.
However, the natural cutoff scale is the Planck scale or
some other high scale, and the associated fine-tuning problem is known as the $\mu$ problem \cite{Kim:1983dt}. In addition
to possibly solving the $\mu$ or other phenomenological problems, the addition of extra matter to the spectrum and/or additional gauge symmetries are
certainly allowed within the context of string theory and often occur in concrete constructions.

\subsection{Singlet-Extended Standard Models and String Theory}
\label{sec:singlet-extended}

A well-motivated class of extensions to the standard model are broadly known as singlet-extended standard models. 
We are concerned here with the supersymmetric versions, which extend
the chiral spectrum of the MSSM by
some number of
MSSM singlets $S_i$. We consider the two specific cases where the exact MSSM spectrum and the MSSM spectrum extended by
three right-handed neutrinos are augmented by the addition of a single $S$ field. Phenomenologically, one reason for adding 
such a field is to generate the $\mu$-term
dynamically via the superpotential term
$
\lambda \,\,SH_uH_d,
$
where the $\mu$-term will be $\mu\sim\lambda \langle S\rangle$ upon $S$ getting a VEV. Since
$S$, $H_u$, and $H_d$ are coupled in the scalar potential, the VEV of $S$ is typically of electroweak
scale up to a few TeV, providing an acceptable $\mu$ if $\lambda$ is not too small. We therefore expect
$\lambda\sim O(1)$\footnote{Other natural
solutions to the $\mu$ problem include the Giudice-Masiero mechanism \cite{Giudice:1988yz} and the direct generation of $\mu$ by string instantons \cite{Blumenhagen:2006xt}.}.

Particular phenomenological models within the class of singlet-extended standard models are broadly classified
by the form of superpotential terms involving $S$, which play a crucial role in phenomenology. We
write the superpotential as
\begin{align}
W = Y_u \,\, q_L  H_u  u_R  + Y_d \,\, q_L H_d d_R + Y_l \,\, L H_d E_R + 
	\lambda \,\, SH_uH_d + W_{\nu} + W_{\text{model}},
\end{align}
which contains the quark and lepton Yukawa couplings of the MSSM superpotential, as well as the term
which generates the $\mu$-term dynamically, some terms $W_\nu$ which account for the neutrino masses, and 
some terms $W_\text{model}$ which are dependent upon the particular singlet-extended model.
The simplest include the nearly minimal
supersymmetric standard model (nMSSM \cite{Panagiotakopoulos:1999ah,Panagiotakopoulos:2000wp,Dedes:2000jp,Menon:2004wv}), the $S^2$  model, and the next-to-minimal supersymmetric standard model\footnote{The problem of cosmological domain walls due to the spontaneously broken $\mathbb{Z}_3$ symmetry
in the original version of the NMSSM \cite{Maniatis:2009re,Ellwanger:2009dp} may be resolved here,
e.g., by the presence of other instanton-induced operators.} (NMSSM \cite{Ellis:1988er}),
which have superpotential contributions of the form\footnote{The coefficients for the models in
(\ref{wmodels}) should be nonzero to avoid an unwanted global symmetry and associated
massless pseudoscalar. The latter is ``eaten'' in the UMSSM. The quartic terms in the NMSSM for $\kappa \ne 0$ (or the $D$ term in the UMSSM) prevent a runaway potential in the $S\rightarrow \infty$,
$H_{u,d}=0$ direction. In the other cases the absence of the runaway direction places constraints on the soft mass-squared term for $S$.}
\begin{align}
\text{nMSSM}:& \qquad W_\text{model} = c_s\,S \notag \\
S^2:& \qquad W_\text{model} = \alpha \,S^2 \label{wmodels} \\
\text{NMSSM}:& \qquad W_\text{model} = \kappa\,S^3\notag.
\end{align}
Another possibility, not considered in this paper, is the UMSSM, which involves
 an extra non-anomalous  $U(1)'$ gauge symmetry \cite{Suematsu:1994qm,Cvetic:1995rj,Cvetic:1997ky}.
Each of these models
have interesting phenomenological implications. 
For example, the extended Higgs sector associated with $S$ may significantly modify the existing and future
searches for the Higgs because of decays into pairs of light pseudoscalars \cite{Chang:2008cw} or because of
reduced couplings due to mixing with the singlet \cite{Barger:2006dh}. Similarly,
the larger neutralino sector may lead to extended cascades at the LHC or
modified cold dark matter scenarios, while the soft cubic supersymmetry breaking
term associated with $\lambda S H_u H_d$ greatly facilitates the possibility of the
strong first-order phase transition needed for electroweak baryogenesis \cite{Accomando:2006ga,Barger:2007ay,Maniatis:2009re,Ellwanger:2009dp}.

While it is clear that string theory can realize singlet-extended standard models at the level
of the spectrum, since singlets so often appear in string constructions, it is interesting to ask more
model-dependent questions related to the presence of certain superpotential terms. If a singlet-extended
standard model is to arise in type II string theory, the singlet $S$ carries global $U(1)$
charge under the branes and thus a coupling of the form $S^n$ is perturbatively forbidden. Fortunately,
a D-instanton which generates these terms non-perturbatively gives corrections of the form
\begin{equation}
\label{eqn:S^n}
e^{-S^{cl}_{E2}}\,\,M_s^{3-n}\,\,S^n \equiv c_n \,\,S^n,
\end{equation}
which allow for many types of singlet-extended models, including the ones described above.
Moreover, it is worth noting that though a single instanton does not have the correct charged
zero mode structure to generate multiple $S^n$ terms with differing $n$, this can be achieved by
superpotential corrections from multiple instantons with different intersection numbers with
the gauge branes.

The presence of a term of the form $S^n$ ensures the absence of a massless pseudoscalar associated
with the Peccei-Quinn symmetry which the term explicitly breaks. In these models the prefactor
$c_n$ must be sufficiently large for $n=1,2,3$ to ensure that the mass of the pseudoscalar
is sufficiently large. One expects the mass to be\footnote{We have assumed that the coefficients of the of the soft supersymmetry breaking terms
corresponding to $c_n S^n$ are $c_n A_n$ where $A_n \sim \langle S \rangle$. For a detailed treatment of the $n=1,3$ cases,
see \cite{Barger:2006dh}.}
\begin{equation}
M_A^2 \sim c_n \langle S\rangle ^{n-1},
\end{equation}
and a reasonable range for the mass is $M_A \sim (3$ GeV $- 1\,$ TeV$)$. Ensuring that the
mass is in this range for $\langle S\rangle= 1\, TeV$ requires
$c_1 \sim (3$ GeV $- 1 $ TeV$)^2$, $c_2 \sim (0.01$ GeV$-1$TeV$)$ and $c_3 \sim (10^{-5} - 1)$. In a
particular quiver, an important question is whether or not the instanton suppression
factors allow for $c_n$'s in these ranges. We will answer this explicitly in our examples.

\subsection{Model-Independent Singlet-Extended Quivers}
Relative to the breadth of possibile particle physics models afforded by string theory,
these three singlet-extended models have far more in common than they do differences, and it
is interesting to study with what frequency and under what conditions these models might arise in
the string landscape. We now investigate this with a detailed quiver analysis at the level of couplings
and will present three singlet-extended quivers, each of which can accomodate the nMSSM, the $S^2$ model, 
the NMSSM, or a more general theory with $W_\text{model}$ a polynomial in $S$.

We performed two systematic analyses of three-stack and four-stack quivers, one which realizes the
exact MSSM chiral spectrum, and one which realizes the exact MSSM extended by three right-handed
neutrinos. Both required stringent string theoretic and phenomenological constraints on the quivers, and all
surviving quivers have non-zero masses for all three families of quarks and leptons, as well as no R-parity violation
or rapid dimension-five proton decay at either the perturbative or non-perturbative level. Additionally,
the quivers have natural mechanisms for realizing the correct order of the top-quark Yukawa coupling and
the neutrino masses. For more details on the methodology and constraints, we refer the reader to
appendix \ref{sec:quiver appendix}.

Interestingly, there are only three\footnote{The scarcity of singlet-extended quivers relative to the $\sim 50$
MSSM quivers of previous work is due to additional
constraints and the investigation here of only three-stack and four-stack quivers. It is not that the MSSM is more common in string constructions.} three-stack or four-stack quivers, given in Table \ref{table:quivers},
 which satisfy all of the string theoretic and phenomenological constraints. All have the exact MSSM spectrum extended by three right-handed neutrinos and a singlet $S$\footnote{We require that the right-handed neutrinos and the singlet $S$ can be distinguished from one another. This is accomplished
by having them transform differently under the D-brane gauge groups and by ensuring the absence of a Dirac-type mass coupling
$LH_uS$. See appendix \ref{sec:quiver appendix} for more details.}. Furthermore, all
exhibit the well-known Madrid embedding \cite{Ibanez:2001nd} of the hypercharge,
\begin{equation}
\label{eqn:madrid}
U(1)_Y = \frac{1}{6} U(1)_a + \frac{1}{2}U(1)_c + \frac{1}{2}U(1)_d,
\end{equation}
where there are four stacks of D-branes $a$, $b$, $c$, and $d$ with $U(3)_a\times U(2)_b\times U(1)_c \times U(1)_d$
gauge symmetry, which becomes $SU(3)_c \times SU(2)_L \times U(1)_Y$ due to the Green-Schwarz mechanism.

In particular, this means that there are \emph{no quivers} with the exact MSSM spectrum plus a singlet $S$ which
satisfy all of the constraints. Two of the quivers with that spectrum nearly pass all constraints, but in those cases
the $\mu$-term is directly generated by an instanton whose presence is necessary to generate an up-flavor
quark Yukawa coupling. As the dynamic generation of the $\mu$-term is a main goal in examining singlet-extended
standard models, we do not consider those quivers.

\begin{center}
\TABLE{
\hspace{-.2cm}
\scalebox{.8}{
\begin{tabular}{|c|c|c|c|c|c|c|c|c|c
		|c|c|c|c|c|}\hline  
		\multirow{2}{*}{Quiver \#}& \multicolumn{2}{c}{$q_L$} & \multicolumn{1}{|c}{$d_R$} & \multicolumn{2}{|c}{$u_R$} & \multicolumn{1}{|c}{$L$} & \multicolumn{3}{|c}{$E_R$}  
		& \multicolumn{2}{|c}{$N_R$} & \multicolumn{1}{|c}{$H_u$} & \multicolumn{1}{|c}{$H_d$}& \multicolumn{1}{|c|}{$S$} \\
		
		&$(a,b)$ & $(a,\ov{b})$ & $(\ov{a},c)$ & $(\ov{a},\ov{c})$
		& $(\ov{a},\ov{d})$ & $(b,\ov{d})$ & $(c,d)$
		& ${\Ysymm}_c $ & ${\Ysymm}_d $ & $\Yasymm_b $ & \scalebox{1}{$\ov{\Yasymm}_b $}
		& $(b,d)$ & $(\ov{b},\ov{c})$
	    & $(c,\ov{d}) $ \\
		\hline\hline			
1&2&1&3&2&1&3&2&1&0&3&0&1&1&1\\ \hline
2&2&1&3&2&1&3&0&2&1&3&0&1&1&1\\ \hline
3&0&3&3&0&3&3&1&0&2&0&3&1&1&1\\ \hline
\end{tabular}}
\caption{The three surviving quivers. Each
quiver has its matter content and transformation behavior specified by a single row, and all have the exact MSSM
spectrum extended by three right-handed neutrinos and a singlet $S$.
\label{table:quivers}
}
}

\end{center}

Since the three quivers which do survive avoid many serious phenomenological pitfalls, it is worth investigating
which particular singlet-extended standard models are possible for these quivers by investigating which of the
terms $S^n$ might be generated by non-perturbative D-instanton effects. A quick look at the quivers gives the global
$U(1)$ charge of such couplings
to be
\begin{equation}
Q_a(S^n) = 0 \qquad Q_b(S^n) = 0 \qquad Q_c(S^n) = n \qquad Q_d(S^n) = -n,
\end{equation}
so that an instanton $E2_n$ which intersects the four gauge D-branes as
\begin{equation}
I_{E2_n,a} = 0 \qquad I_{E2_n,b} = 0 \qquad I_{E2_n,c} = n \qquad I_{E2_n,d} = -n
\end{equation}
will generate an $S^n$ coupling of the form \eqref{eqn:S^n}, with a different suppression factor associated with
each instanton. Though this ensures that it is possible to generate the superpotential terms associated with the
nMSSM, the $S^2$ model, or the NMSSM, it does not ensure that the instanton suppression factor, which depends
on the volume of the three-cycle which the instanton wraps, is such that
they are of an allowed order.

In fact, one might question whether it is even possible at the level of quivers to address this issue. Certainly
it is not possible to say what the volume of the instanton cycle is from geometric specifics and stabilization
arguments, since we have not specified any geometry, but it turns out that the issue can 
be addressed phenomenologically. For example, it was pointed out in \cite{Ibanez:2008my} that in some MSSM quivers,
the instanton which generates the $\mu$-term is also required to generate a quark or lepton Yukawa coupling. To generate
a Yukawa coupling of the correct order would require the suppression factor $e^{-S^{cl}_{E2}}\gtrsim 10^{-5}$, which is
not nearly suppressed enough to generate a $\mu$-term $e^{-S^{cl}_{E2}} M_s \, H_uH_d$ of the correct order, and the quiver must
be phenomenologically ruled out\footnote{Such a quiver would be possible for a lower
 string scale, $M_s\lesssim 10^7$ GeV. See \cite{Cvetic:2010qy} for an 
analysis of related issues in type II orientifold compactifications.}.

A similar analysis of the charged zero mode structure of instantons which
generate an $S^n$ coupling shows that only one of these instantons might also generate
a quark or lepton Yukawa coupling. Specifically, the instanton $E2_1$ which generates
a superpotential term linear in $S$ will generate a charged lepton Yukawa coupling $LH_dE_R$ in either
of the first two quivers in Table \ref{table:quivers}. For phenomenological reasons the $c_s$ parameter of the nMSSM
 must be $O({\rm TeV}^2)$, though, which requires a suppression factor $e^{-S^{cl}_{E2}} \sim 10^{-30}$, and
thus $E2_1$ cannot simultaneously account for the correct order of the charged lepton Yukawa coupling and $c_s$.
However, this same lepton coupling might also receive a contribution from an instanton with
an extra pair of charged vector-like zero modes $\lambda_c$ and $\ov{\lambda}_c$\footnote{See \cite{Cvetic:2009ez} for
more details on instantons with vector-like zero modes in quiver analyses.}. If such an instanton with
vector-like zero modes accounts for the correct order of the lepton Yukawa coupling, then the suppression
factor associated with $E2_1$ is not fixed, and $E2_1$ could generate a superpotential term linear
in $S$ with any suppression factor. Furthermore, based on global $U(1)$ charges, one might be concerned that
the instanton which generates an $S^n$ coupling also generates an R-parity violating of the form $(q_L L d_R)^n$.
Such a term will generically be suppressed by $a_n/M_s^{2n}$, where the desirable superpotential coupling 
takes the form $a_nS^n$, and thus is not dangerous.

	 These three singlet-extended quivers are model independent,
as any superpotential term $S^n$ can be generated non-perturbatively by a D-instanton, and can thus realize
the nMSSM, the $S^2$ model, the NMSSM, or any other theory with $W_\text{model}=f(S)$, with $f$ a polynomial. 
Moreover,
the corresponding instanton suppression factors are not phenomenologically fixed. Thus, the quivers serve as an
excellent starting point for building global type II orientifolds with the spectrum of a singlet-extended
standard model. Determination of which particular model is embodied by such 
a global type II orientifold would then be possible, as one could explicitly determine the form of instanton
corrections to the superpotential via geometric and CFT techniques.

	In addition to allowing one to determine the explicit superpotential corrections due to instantons, such a global
construction also allows one to address in detail the values of the instanton suppression factors. This is
accomplished by examining the details of K\" ahler moduli stabilization, where the vacuum expectation value
of the real part of an appropriate K\" ahler modulus is the volume of the cycle which the instanton wraps,
which is the crucial parameter that determines the instanton suppression factor. While it is not possible at
the quiver level to state the precise value of the instanton suppression factors, since a global construction has
not been specified and the K\" ahler moduli have not been stabilized, the ranges for $c_{1,2,3}$ discussed
in section \ref{sec:singlet-extended} which allow for a light pseudoscalar 
are certainly achievable due to the exponential supression. Alternatively, if the cycle which the instanton
wraps is stabilized at a relatively small volume, $c_n$ could be large and allow for a high mass pseudoscalar.
This is in contrast to the heterotic scenario studied in \cite{Lebedev:2009ag}, where the vacuum expectation values
of standard model singlets which determine $c_3$ typically give $c_3\ll 1$, and thus give a light pseudoscalar.

Finally, since these quivers exhibit many other nice phenomenological features\footnote{Beyond the requirement
that the top mass be very massive relative to other MSSM matter fields, we have not addressed mass hierarchies or mixing angles
in this analysis. The first two quivers, in particular, would require some amount of fine-tuning of worldsheet instantons
to obtain realistic mixing angles.}, it is worth commenting on
the possible forms which $W_\nu$ might take for each quiver in order to account for the observed order of the
neutrino masses. In the first two quivers the Dirac neutrino mass term $LH_uN_R$ can be generated by two instantons,
which we call $E2$ and $E2_v$, where the latter has an additional pair of vector-like charged
zero modes $\lambda_d$ and $\ov \lambda_d$. The observed order of the neutrino masses can be accounted for if
$E2_v$ generates a highly suppressed Dirac mass term. However, this is not true of $E2$, since its presence
will also generically generate a Majorana mass term $N_RN_R$, giving rise to the seesaw mechanism. With
$M_s \approx 10^{18}$ GeV, the associated seesaw mass would give a contribution to the neutrino mass a few
orders of magnitude below what is observed, and thus for the seesaw mechanism to work at high string scale, $E2_v$
must also generate a contribution to the Dirac mass term\footnote{This issue of the same instanton generating
the Dirac mass term and the Majorana term was discussed in depth in \cite{Cvetic:2010qy}, where it was shown that the
string scale must be lowered for the corresponding seesaw mass to generate neutrino masses of the observed order. }.
In the third quiver, the Dirac mass is perturbatively allowed, and therefore the option of a highly suppressed 
non-perturbative Dirac mass is not available. The seesaw mechanism can be realized, however, if a Majorana mass term
is generated by an instanton.

\section{Conclusions}

String theory naturally gives rise to two of the most important features of experimental particle physics, 
namely gauge symmetry and chiral matter. Recently, however, improved knowledge of non-perturbative
effects which are present in string models has been used to explain finer details of particle physics, 
such as hierarchical Yukawa couplings. 
This progress has led to a great deal of work in bottom-up string 
model building in the context of type II orientifold compactifications.
However, it is also important to study the MSSM and its most likely extensions from the perspective
of the string landscape.

Of great use has been the notion of a quiver, 
which is a subset of the data associated with a string model that allows for the efficient 
investigation of physical effects at the level of couplings. Systematic phenomenological studies
of MSSM quivers have been performed which investigate the role of D-instantons in generating quark and
lepton Yukawa couplings while still avoiding phenomenological pitfalls, such as R-parity violating
couplings and rapid dimension-five proton decay.

  In this work, we moved beyond the MSSM and investigated the class of singlet-extended
  (supersymmetric)  standard models, which add
a singlet $S$ to the MSSM spectrum, possibly also extended by three right-handed neutrinos. Phenomenologically,
the motivations include the dynamical generation of the $\mu$-term by a coupling $SH_uH_d$ when
$S$ gets a vacuum expectation value, as well as interesting possibilities for extended Higgs and neutralino
sectors, with implications for the LHC, cold dark matter, and electroweak baryogenesis.
 Such extensions are also motivated by string models, which often contain one or
many MSSM singlets. The nMSSM, $S^2$ model, and NMSSM are notable examples of singlet-extended standard models,
with the difference between the three being the structure of superpotential terms involving the $S$ field.

It was the goal of this paper to efficiently map out regions of the landscape which might allow
for singlet-extended models, taking into account non-perturbative effects. In section \ref{sec:type II},
we reviewed the basics of quivers in type II orientifold compactifications as well as non-perturbative
superpotential corrections due to D-instantons. In section \ref{sec:beyond} we motivated the class of 
singlet-extended standard models and discussed the role of instanton effects in generating superpotential
terms of the form $S^n$.

We then presented
the three four-stack singlet-extended quivers which satisfy all of the string theoretic and phenomenological constraints
in appendix \ref{sec:quiver appendix}, which include the presence of non-zero masses for all three families
of quarks and leptons and of a perturbative $S H_u H_d$ term,  and the absence of R-parity violating couplings and rapid dimension-five proton decay
at both the perturbative and non-perturbative level. All three of these quivers exhibit the spectrum of the
MSSM extended by three right-handed neutrinos and a singlet $S$ and furthermore have the Madrid hypercharge embedding. Moreover,
all three quivers generically allow for the non-perturbative generation of a superpotential of the form $f(S)$,
where $f$ is a polynomial. In this case, each monomial would be generated by a different instanton with
a suppression factor that is not phenomenologically fixed, and thus the quivers are completely model independent
at the level of the superpotential. Thus, they can in principle realize the nMSSM, the $S^2$ model, and the NMSSM. Furthermore, all have mechanisms for small neutrino masses.

This work has shown the efficiency of the quiver approach in identifying promising patches of the string landscape.
It would be interesting to construct a global type II orientifold compactification which realizes one of these quivers,
which would allow for the classification of instanton corrections to the superpotential in that particular
string model and thus the explicit determination of $f(S)$ in that model. It would also be interesting to
conduct similar studies of the quiver landscape using other promising phenomenological models,
such as singlet-extended models involving an additional non-anomalous $U(1)'$ gauge symmetry \cite{Suematsu:1994qm,Cvetic:1995rj,Cvetic:1997ky}.
\vspace{2cm}
\section*{Acknowledgements}
We acknowledge useful conversations with Robert Richter and are grateful for his participation in
past collaborations.
We acknowledge the hospitality 
of the KITP during the Strings at the LHC and in the Early Universe program for
providing a stimulating environment during the completion of this work.
This research was supported in part by the National Science Foundation under Grant No. PHY05-51164.
The work of M.C. and J.H. is supported by the DOE Grant DOE-EY-76-02-3071, the NSF RTG grant DMS-0636606, the Fay R. and Eugene L. Langberg Chair and in part by the Slovenian Research
Agency (ARRS). The work of P.L. is supported by the IBM Einstein Fellowship and by the NSF grant PHY-0503584.
\appendix
\section{The Systematic Analysis: Methodology and Constraints\label{sec:quiver appendix}}
Having motivated the use of a quiver analysis in identifying promising vacua in the string landscape, in this
appendix we briefly describe the methodology and constraints employed in the systematic search which produced the
quivers presented in the main text.
While the quivers are not specific to type II string theory at the level of gauge symmetry and chiral matter, the phenomenological
analysis strongly relies on the presence of global $U(1)$ symmetries, possible non-perturbative superpotential corrections due to
D-instanton effects, and a $U(1)_Y$ gauge symmetry which is not given a mass by the generalized Green-Schwarz mechanism. These
ideas naturally arise in type II, motivating the use of its language rather than the language of nodes and arrows. The map
is straightforward, though, as a stack of D-branes give rise to a node, and an open string state between two stacks of intersecting
D-branes gives rise to an arrow between two nodes.

\vspace{.3cm}
\noindent\textbf{Methodology}
\vspace{.3cm}

\noindent A systematic phenomenological analysis of the type II quiver landscape begins with a reasonably small set of input data.
First, one must decide which phenomenological theory is of interest and choose the number of stacks with which to realize the
gauge symmetry.
For example, in the MSSM one needs a stack of three and a stack of two D-branes\footnote{Unless the $SU(2)_L$ factor
of the MSSM is realized as an $Sp(2)\cong SU(2)$ gauge symmetry by a single D-brane on a three-cycle homologically invariant
under the orientifold action.} to give the requisite $SU(3)_c$ and $SU(2)_L$, but one might also add many stacks which contain a single
D-brane. We chose to examine the cases of one or two additional stacks with a single D-brane, giving a total of three stacks and
four stacks.

In addition to the specification of the number and types of D-brane stacks, which gives the gauge symmetry of the theory,
one must also specify the chiral matter spectrum.
In the work of \cite{Cvetic:2009ez,Cvetic:2009ng,Cvetic:2009yh,Cvetic:2010qy}, the authors considered the exact MSSM or the
exact MSSM extended by three right-handed neutrinos. In this work we considered these two sets of spectra extended by an
MSSM singlet
$S$ which should not be interpreted as a right-handed neutrino.

As the MSSM and its extensions have a hypercharge $U(1)_Y$ gauge symmetry and the $U(1)$ gauge symmetries associated with a stack of D-branes are
generically lifted by the generalized Green-Schwarz mechanism, it is also necessary to specify the linear combination $U(1)_Y = \sum q_xU(1)_x$
which is left massless. It turns out that the chiral matter spectrum along with necessary constraints for tadpole cancellation and a
massless hypercharge put strong restrictions on the allowed linear combinations, often giving a finite number of
combinations which can realize the hypercharge. This was the case in the analyses performed in this paper.

Given the number and type of D-brane stacks, the chiral matter spectrum of interest, and the massless linear combination of $U(1)$'s
which is interpreted as hypercharge, it is possible to write down every way in which a given matter field might transform. For example,
in the Madrid embedding,
in \eqref{eqn:madrid},
the right-handed down-quarks $d_R$ might be realized as $\Yasymm_a$, $(\ov{a},c)$, or $(\ov{a},d)$. Given that there are three families
and the fact that different families might arise in different representations of the D-brane gauge groups, it is possible to
enumerate all possible ways in which three $d_R$'s might transform. One could do this for every field in the chiral
spectrum, which allows for the straightforward enumeration of all possible quivers with this number of stacks, matter
spectrum, and hypercharge embedding.

Most of these quivers do not satisfy necessary string theoretic constraints for tadpole cancellation
 and a massless hypercharge. Additionally, quivers which do satisfy those constraints often
exhibit undesirable phenomenological effects.
For these reasons, we enforce extensive constraints which ensures that the surviving quivers are
theoretically and phenomenologically viable.

\vspace{.3cm}
\noindent\textbf{Theoretical and Phenomenological Constraints}
\vspace{.3cm}

In this work we performed two systematic analyses, with the difference being the presence or absence
of right-handed neutrinos in the spectrum. This difference motivates slightly different
phenomenological constraints, so we first discuss the constraints common to both analyses and then
discuss the additional constraints placed on the analysis of quivers which exhibit right-handed
neutrinos. We briefly discuss constraints which were used in previous analyses, referring the reader to \cite{Cvetic:2009ez,Cvetic:2009ng,Cvetic:2009yh,Cvetic:2010qy}
for more details, and discuss
new constraints related to the presence of a singlet $S$ in the
spectrum.

As mentioned in section \ref{sec:type II}, the cancellation of Ramond-Ramond charge in the Calabi-Yau,
also known as tadpole cancellation, places constraints on the homology of the D-branes and O-planes. There
are also additional constraints on the homology of the D-branes and O-planes if a linear combination
of $U(1)$'s is to be left massless by the Green-Schwarz mechanism. Since the homology of D-branes determines
the chiral matter spectrum of the theory, these two constraints each place a constraint on the chiral
matter spectrum which \emph{must} be satisfied if the quiver is to cancel tadpoles and have a massless
hypercharge when embedded in a top-down string model. In previous work, these have been called ``top-down" constraints
not because top-down globally consistent string models were presented, but because the constraints on the chiral matter 
arise from string theory. The constraints on the chiral matter which arise from string theory are equivalent
to non-abelian and abelian anomaly cancellation in the low energy effective theory, so they can also be viewed
as bottom-up. 

In addition, we require
that the quivers satisfy many phenomenological constraints. First, since many quark and lepton Yukawa couplings
are perturbatively forbidden, we require that enough of these forbidden couplings are non-perturbatively generated
by D-instantons to ensure non-zero masses for all three families of quarks and leptons. However, these same instantons which are
required to generate Yukawa couplings might also generate phenomenological drawbacks. Therefore, we require the
absence of the R-parity violating couplings $d_Rd_Ru_R$, $LLE_R$, $q_LLd_R$, and $LH_u$ and the absence of the dimension-five proton decay operators $q_Lq_Lq_LL$ and $u_Ru_Rd_RE_R$ on both the perturbative
and non-perturbative level. We also require that there is a natural explanation for the size of the top-quark
Yukawa coupling. These constraints were also present in systematic analyses presented in previous work.

We further require that three new constraints are satisfied which maintain the motivations for looking at
singlet-extended standard models. First, since we wish to have a dynamical $\mu$-term, we
require that the coupling $H_uH_d$ is absent at both the perturbative and non-perturbative levels, while $SH_uH_d$ is perturbatively
realized. In addition, we require that the Dirac-type coupling $LH_uS$ is absent at both 
levels, so that the singlet $S$ should not be interpreted as a right-handed neutrino.
This allows us to isolate the issues of
neutrino mass and the scale of the $\mu$-term. Finally, we require that either a linear, quadratic, or cubic
term in $S$ can be generated in the superpotential without giving rise to the phenomenological drawbacks mentioned
in the previous paragraph.

In addition, for the analysis which considers quivers with three right-handed neutrinos,
we require that a linear term in $N_R$ is not generated by an instanton whose presence is required to generate
a forbidden Yukawa coupling. We also require that $S$ and $N_R$ are realized in different D-brane sectors and thus do not
transform in the same way, ensuring the $S$ is distinguishable from $N_R$.
		
\section{Stabilization and Decoupling of Charged Moduli \label{sec:moduli appendix}}
In this appendix we address the issue of the stabilization and decoupling of
charged moduli $C_i$. Given the specific assumption that the stabilization of uncharged moduli $U_i$ (and supersymmetry breaking) takes place at a scale $\gg$ O(TeV), we  deduce that  the real part of charged moduli is  subsequently  determined by the D-flatness conditions while the  imaginary part  of charged moduli is fixed at the TeV scale due to the D-instanton induced couplings to the charged matter $\Phi_i$. For simplicity, we shall denote the collection of charged moduli, uncharged moduli and matter fields with the single set of letters $C$, $U$ and $\Phi$, respectively.

In the type II context charged moduli arise in non-perturbative corrections to the
superpotential from D-instantons which have chiral intersections with one or more
gauge D-branes. For example, recall that in type IIa the superpotential correction
generated by an instanton wrapping a three-cycle $\Xi$ is suppressed by \cite{Blumenhagen:2006xt}

\begin{equation}
e^{-S^{cl}_{E2}}=exp[\,\,\frac{2\pi}{l^3_s}\left(\frac{1}{g_s} Vol_{\Xi} - i
\int_{\Xi} C_{3} \right)\,\,].
\end{equation}
If $\Xi$ intersects a D6-brane wrapping $\Pi_a$, the Ramond-Ramond three-form $C_3$ participates
in the Chern-Simons couplings which cancel anomalies associated with $U(1)_a$ via the generalized
Green-Schwarz mechanism. The corresponding transformation behavior of the three-form gives the
transformation behavior

\begin{equation}
e^{-i\int_\Xi C_3} \mapsto e^{iQ_a(E2)\Lambda_a}e^{-i\int_\Xi C_3},
\end{equation}
where $\Lambda_a$ parameterizes the transformation of the $U(1)_a$ gauge boson $A_\mu$ via 
$A_\mu \mapsto A_\mu +\partial_\mu\Lambda_a$, and $Q_a(E2)= N_a \,\, \Xi \cdot (\Pi_a-\Pi_a')$. In this sense, we refer to the corresponding modulus as a charged modulus. For explicit moduli dependence of three-cycle volumes in the toroidal
type IIa context see, e.g., \cite{Cvetic:2003yd}.

For the sake of simplicity, here we parametrize the instanton suppression factor associated with a
non-perturbative superpotential correction as $e^{-aU-bC}$, where $U$ parameterizes  uncharged  complex structure moduli and $C$
is a specific charged modulus associated with the  $U(1)_a$ factor of gauge branes $D_a$ wrapping a  specific three-cycle $\Pi_a$. 
In the type IIa context the uncharged part of the  non-perturbative coupling parameterizes the complex structure and dilaton moduli associated with those three-cycles which do not intersect with the specific three-cycle $\Pi_a$.

For specificity we assume that the uncharged moduli are stabilized at a scale $\gg$ TeV, with 
 the superpotential and K\" ahler potential  split as 
\begin{equation}
W = W_0(\Phi,U,C) + \tilde W(U)\qquad \qquad 
K = K_0(U,U^*,C+C^*+Q_aV,\Phi,\Phi^*) + \tilde K (U,U^*),
\end{equation}
where $\Phi$ is a chiral field charged under the gauge symmetry of the D-brane with 
anomalous $U(1)_a$. The vector super-multiplet $V$, associated with $U(1)_a$,  gets a  St\"uckelberg mass of  the order ${\tilde M}_s \sim M_{s}$ via the Green-Schwarz mechanism, i.e. $M_V^2\equiv K_{C,C*}|_{V=0}$, and furthermore there is the charged modulus dependent Fayet-Iliopoulos  D-term: $\xi\equiv K_C|_{V=0}$.  Note that  for the sake of specificity we assumed that  the VEV's of the uncharged moduli $U={\cal O}(1)$, which leads to $M_{pl}\sim M_{s}\sim {\tilde M}_s$.  Generalizations to lower string scales are straightforward, but they require a more careful treatment of ${\tilde M}_s$.

$W_0$ and $K_0$ are the low energy superpotential and K\" ahler potential, 
and $\tilde W$ and $\tilde K$ are contributions to the superpotential and K\" ahler potential
from the ``hidden sector'', responsible for the stabilization of the uncharged moduli $U$. 
Again, we assume  that contributions from $\tilde K$ and $\tilde W$ are at a
high (string)  scale  $\tilde M_s\gg$ TeV and are sufficient to stabilize the uncharged moduli $U$ and
break supersymmetry without giving rise to a cosmological constant. In $K_0$, $W_0$ and  the potential $V_0$, then,
$U$'s  are  not  dynamical variables and are replaced  with their non-zero VEV's. This above splitting of $W$ and $K$
is justified when $\langle C\rangle \ll \langle U\rangle $ and $\frac{\langle\Phi \rangle}{\tilde M_s} \ll \langle U \rangle $, which for $\Phi$ is justified since we expect such a chiral
matter field to be associated with low energies, and  the solution for  $C$  is justified  a posteriori by the self-consistency of 
the derived solution as follows.

We examine the possible existence of a consistent solution where ${\rm Re} \  (C)$ and $\Phi$ stabilize near zero VEV. Expanding
the K\" ahler potential around zero VEV in ${\rm Re} \ (C)$ and $\Phi$ and keeping only the quadratic terms, we write\begin{equation}
K_0 = \frac{\tilde M_s^2}{2}(C+C^*+Q_aV)^2 + d \, \Phi\Phi^*, 
\end{equation}
where $d$ is a dimensionless $O(1)$ parameter. Note again, that in the type IIa context ${\rm Re}(C)$ is a complex structure modulus which parametrizes the deviation away from  the special 
Lagrangian cycle wrapped by the $D_a$  brane stack. In the  type IIb context these moduli specify the size of the blow-up moduli for the $D_a$ brane stack wrapping an orbifold singularity. Note that this structure of the K\"ahler potential, which is a quadratic function of $2\, {\rm Re}(C) = C+C^*$, is expanded around zero VEV for $2\, {\rm Re}(C)$. This is specific to charged moduli associated with gauge $D_a$-brane  sectors, and it should be contrasted with those of the gauge coupling modulus in the heterotic string context, which typically have a logarithmic dependence on the dilaton modulus. Studies of moduli stabilization  and supersymmetry breaking with charged moduli in the latter context  can therefore lead to different conclusions. See, e.g. 
\cite{Dudas:2008qf},
  and references therein.  Again,  the mass and Fayet-Iliopoulos parameter associated with the anomalous $U(1)_a$
symmetry are $M_{V}^2\equiv K_{CC^*}|_{V=0}=\tilde M_s^2$ and  $\xi\equiv K_C|_{V=0}=\tilde M_s^2\,(C+C^*)$, respectively.

The effective low energy potential   can now be  written as a sum of the low energy $\cN=1$ global scalar potential for $\Phi$   fields and their soft  supersymmetry breaking terms (from
the hidden sector $\tilde W(U)$),  as well as a potential F-term  contribution due to charged moduli $C$, and we write these contributions as $V_0$. Furthermore, there is the $D$-term contribution of
the anomalous $U(1)_a$, so that the scalar potential is of the form
\begin{equation}
V\,\,=\,\,V_0(\Phi,\langle U\rangle,C)+\frac{g^2}{2}D^2\,\,=\,\,V_0(\Phi,\langle U\rangle,C) + \frac{g^2}{2}\tilde M_s^4(C+C^*-\frac{Q_a|\Phi |^2}{\tilde M_s^2})^2.
\end{equation}
Minimizing this potential with respect to $C$, we see that
\begin{equation}
0=\frac{\partial V}{\partial C} = \frac{\partial V_0}{\partial C} + g^2 \tilde M_s^4(C+C^*-\frac{Q_a|\Phi |^2}{\tilde M_s^2}),
\end{equation}
which gives
\begin{equation}
\label{eqn:stabilized C}
C+C^* = \frac{Q_a|\Phi |^2}{\tilde M_s^2} - \frac{1}{g^2\tilde M_s^4}\frac{\partial V_0}{\partial C} =  O(\frac{{\rm TeV}^2}{\tilde M_s^2}) + O(\frac{{\rm TeV}^4}{\tilde M_s^4})\approx 0,
\end{equation}
and thus $C+C^*$ is stabilized very close to zero. Note that the second term is suppressed by ${\rm TeV}^2/\tilde M_s^2$ relative
to the first and can be therefore be dropped. Integrating out the charged modulus and minimizing the potential 
with respect to $\Phi$, we obtain
\begin{align}
\label{eqn:b8}
\frac{\partial V}{\partial \Phi} &= \frac{\partial V_0}{\partial \Phi} + g^2\tilde M_s^4(\frac{-Q_a\Phi^*}{\tilde M_s^2})(C+C^*-\frac{Q_a|\Phi |^2}{\tilde M_s^2}) \notag \\
&= \frac{\partial V_0}{\partial \Phi} + g^2\tilde M_s^4(\frac{-Q_a\Phi^*}{\tilde M_s^2})(- \frac{1}{g^2\tilde M_s^4}\frac{\partial V_0}{\partial C})
\sim O({\rm TeV}^3) + O(\frac{{\rm TeV}^5}{\tilde M_s^2}).
\end{align}
Again, the second term is suppressed by a factor of TeV$^2/\tilde M_s^2$ relative to the first. Moreover,
minimization with respect to $C$ forces the $D$-term contribution to vanish, up to contributions of $O({\rm TeV}^8/\tilde M_s^4)$. This
is crucial, since by examining two terms in the $F$-term contributions to the scalar potential, we see
\begin{equation}
\label{eqn:b9}
V_0(\Phi,\langle T\rangle,C) \sim \frac{1}{K_{\Phi\Phi^*}}|D_\Phi W_0|^2 + \frac{1}{K_{CC^*}}|D_c W_0|^2\sim O({\rm TeV}^4) + O(\frac{{\rm TeV}^6}{\tilde M_s^2}).
\end{equation}
Note again, that the D-term  contribution to the scalar potential is suppressed  by a factor of TeV$^4/{\tilde M_s^4}$ relative
to the first term in \eqref{eqn:b9}, and can therefore be dropped relative to $V_0$. Furthermore, from \eqref{eqn:b9} we  see that the second term (that involves F-terms due to charged moduli)  is suppressed
by a factor of ${\rm TeV}^2/\tilde M_s^2$ relative to the first, and it can also be dropped.

In conclusion, the real part of the charged modulus $C$ is stabilized near zero VEV due to  the D-flatness constraint, or equivalently, 
integrating  out  the heavy vector super-multiplet  enforces its  $D$-term to effectively vanish. Moreover, terms involving $C$ in the
$F$-term contribution to the scalar potential   are  highly suppressed, 
i.e.  thus  the only contribution to $V_0$ arises by replacing ${\rm Re}(C)$ by its zero VEV. \begin{equation}
V_\text{eff} = V_0(\Phi,\langle U\rangle, {\rm Re} (C)\approx 0).
\end{equation}
Thus, the  dynamics of the charged matter fields $\Phi$ is determined by its $\cN=1$ global potential,  soft supersymmetry breaking terms  terms arising from the hidden sector stabilization
of the uncharged modulus $U$, and replacing  the real part of ${\rm Re} (C)$ with near-zero VEV. Note, however, that the low energy potential can still depend on ${\rm Im} (C)$, due to soft supersymmetry breaking terms, and will typically
receive a mass of the order  of the TeV scale.

Given the assumption of $U$ stabilization in a hidden sector
which breaks supersymmetry and leaves zero cosmological constant,  the stabilization  scenario of the real part of the
charged modulus is quite general. 
We point out  that the analysis and conclusions  spelled out above  are closely related to those  of
\cite{GarciadelMoral:2005js}.
We have, however, allowed for general soft supersymmetry breaking terms, which could lead to nonzero VEV's of charged matter fields $\Phi$ at the TeV scale. We have also carried out the  study of the  charged moduli decoupling  in the presence of non-perturbative, charged modulus
 dependent superpotential couplings.

\bibliographystyle{JHEP}
\bibliography{refs}

\end{document}